\documentclass[12pt,a4paper]{article}
\usepackage{epsfig}
\def\eq#1{{Eq.~(\ref{#1})}}
\newcommand{\Le}{\left(}
\newcommand{\Ra}{\right)}

\newcommand{\beq}{\begin{equation}}
\newcommand{\eeq}{\end{equation}}   
\newcommand{\beqar}[1]{\begin{eqnarray}\label{#1}}
\newcommand{\eeqar}{\end{eqnarray}}
\begin{document}
\title {{\Huge \bf An Inclusive Cross Section for the }\\
{\Huge \bf Nucleus - Nucleus Interaction}\\
{\Huge \bf at RHIC Energies}}
\author{
{\Large\bf S. ~B o n d a r e n k o,\thanks{Email: serg@post.tau.ac.il}
\quad
E.~G o t s m a n,\thanks{E-mail: gotsman@post.tau.ac.il}}\\
{\Large \bf 
\quad  E.~L e v i n\thanks{E-mail: leving@post.tau.ac.il} \quad and \quad
\quad U.M a o r\thanks{E-mail: maor@post.tau.ac.il}
} \\[10mm]
{\it\normalsize HEP Department}\\
 {\it\normalsize School of Physics and Astronomy,}\\
 {\it\normalsize Raymond and Beverly Sackler Faculty of Exact Science,}\\
 {\it\normalsize Tel-Aviv University, Ramat Aviv, 69978, Israel}\\[0.5cm]
}

\date{January 8, 2001}

\maketitle
\thispagestyle{empty}

\begin{abstract} 
 We discuss the  saturation of the parton density in heavy
ion
collisions at RHIC energies
using a  Pomeron approach. Our predictions for the
particle
density in ion-ion collisions at RHIC energies can be utilized as 
the background for the  observation of  possible quark-gluon plasma
production.
\end{abstract}
\thispagestyle{empty}
\begin{flushright}
\vspace{-16.5cm}
TAUP 2663-2001 \\
January 8, 2001\\
{\tt hep-ph/0101060}\\           
\end{flushright}

\newpage

\section*{}
For heavy ion collisions at high energy,  saturation of the
parton density is predicted 
both in  high density QCD \cite{GLR,MUQI,MLVE} and
in a  Pomeron approach \cite{AA}. 
The goal of this letter is to estimate the value of
the inclusive cross section for
the case of nucleus-nucleus interactions in a  Pomeron approach, as this
will be the
 background for any interesting signal of new physics such as
the  saturation of the gluon density \cite{GLR,MUQI,MLVE}, or  the 
creation
of a
quark-gluon plasma at  RHIC \cite{Bass}.
The physical picture,  behind our
Pomeron approach, is the old fashioned parton model \cite{PARTONS}, which
was and can still be used as a guide in  strong interaction physics
at large
distances ("soft" physics).
In spite of the simple physics described by Pomeron exchanges, the Pomeron
approach is  technically rather  complicated, and this problem has not
yet been 
solved for hadron-hadron interactions. However, for inclusive production
in ion-ion collisions there are two significant simplifications which
enable one to develop a self consistent theoretical description
in our  Pomeron approach: 

\begin{enumerate}
\item\,\,\, For a heavy nucleus the vertex of the Pomeron-nucleus
interaction $
g_{PA}(b_t) = A^{\frac{1}{3}}\,\,g_{PN}(b_t = 0 ) \,S_A(b_t)$ where 
$g_{PN}(b_t = 0)$, is the vertex of the Pomeron - nucleon interaction at
$b_t = 0$,
and $S_A(b_t)$ is the nucleus profile function normalized so that $\int
\,\,d^2 b_t\,\,S_A(b_t) = 1 $.    Due to $A^{\frac{1}{3}}$ enhancement of
Pomeron-nucleus interaction we can develop a theoretical 
approach  \cite{SCHWIM} 
considering
$G_{3P}\, g_{PA}(b_t=0 ) \,\approx\,1$ while $ G^2_{3P}
\,\,\ll\,\,1$. Here, $G_{3P} $ is the triple Pomeron vertex (see Fig. 1 ). 
In this approach one  can neglect the loop Pomeron diagrams in comparison
with the "tree" diagrams of  Fig.1.

\item\,\,\,Using the AGK-cutting rules \cite{AGK}, 
all Pomeron diagrams in
which more then two Pomerons cross the rapidity level $y_c$  cancel in
the inclusive
cross section. Therefore, only Mueller diagrams, shown    in Fig.1,
survive.
\end{enumerate}

Using the approach, developed in Refs.\cite{SCHWIM,AA,JEN}, we obtain 
the following closed expression for the diagrams shown in Fig.1 ( see
Fig.1 for notation )

\begin{figure}[htbp]
\begin{center}
\epsfig{file=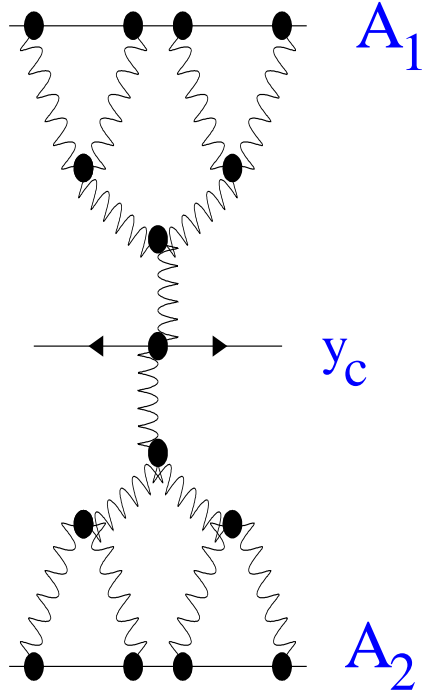,width=6cm}
\end{center}
\caption{\it The Mueller diagram for an inclusive   A-A
interaction process.}
\label{fig1}
\end{figure}

\begin{eqnarray}
\frac{d \sigma (A_1 + A_2 )}{d y_c} &= &\int \,\,d^2 b_t \,\,d^2 b'_t
\,\,a^P_P\,\,R_{A_1}\Le Y - y_c, b_t \Ra \,\cdot\,R_{A_2 }\Le y_c, b'_t
\Ra =
\label{E1} \\
 &=&  a^P_P\,\,\int \,\,d^2 b_t \,\,d^2 b'_t
\,\,\frac{g_{PA_1}(b_t)\,g_{PA_2}(b'_t)\,\, e^{\Delta Y}}{ \Le
\, g_{PA_1}(b_t)\,\gamma\, \Le e^{\Delta\,(Y - y_{c})}-1 \Ra \,\,+
\,\,1\,\Ra \, \, \Le g_{PA_2}(b'_t)\,\gamma\,\Le
e^{\Delta \,y_c}-1\Ra \,\,+\,\,1\,\Ra}\,\,\nonumber
\end{eqnarray}
here $\gamma=G_{3P}/\Delta$ and  $G_{3P}$ is the vertex of the triple
Pomeron interaction, $\Delta = \alpha_P(0) - 1$ where $ \alpha_P(0)$  is
the Pomeron's intercept,
$a^P_P$ is the particle emission vertex and
$Y=\ln\Le s/1\,GeV^2\Ra$ is the  total rapidity interval  for $A_1 - A_2$ 
collision. $R_A(Y,b_t)$ is the sum of "tree" diagrams 
and is given in \cite{SCHWIM,AA}. 

  Eq.(~\ref{E1})  predicts  saturation of the density of produced
particles
which can be defined as 
\beq \label{E2}
\rho \left( y_{c}\right)\,\,\,=\,\,\,\frac{d  \,N(y_c)}{d
\,y_c}\,\,\,=\,\,\,
\frac{\frac{d \sigma (A_1 + A_2 )}{d y_c}}{\sigma_{tot}( A_1 + A_2)},
\eeq
where $N(y_c)$ denotes the multiplicity of particles with rapidity $y_c$.

   Eq.(~\ref{E1}) leads to a constant  density  $\rho(y_c)$ at high
energies.    To show this, we  note that using a simplified Wood-Saxon
type  $b_t$ distribution for the  density  of a nucleus, namely,
\beq\label{I5}  
S_A\Le b_t\Ra=\frac{1}{\pi\, R^2_{A}}\,\Theta\Le R_{A}-b_t\Ra\,\,,
\eeq
the total cross section ( for $ A_1 \,>\,A_2$ ) is 
\beq
\sigma_{tot}=2\,\pi\,R_{A_1}^{2}.
\label{I6}
\eeq
Integrating over $b_t$ and $b'_t$ in Eq.(~\ref{E1}) we  have
\beq \label{E3}
\rho(y_c)\,
\,\longrightarrow|_{Y \gg y_c \gg 1}
\,\,\,\,\,a^P_P\,\,\frac{ R^2_{A_2}}{2 \pi\,\, \gamma^2}\,\,,
\eeq
where at high energy (see Fig.1)
$$
a^P_P=\frac{\frac{d \sigma (N + N )}{d
y_c}}{\sigma_{tot}( N + N)}
$$
 Eq.(~\ref{E3}) suggests that the particle density   
is independent of  the energy and of the atomic number of the heaviest
nucleus.
It is instructive to compare this prediction with the density obtained in
the
Glauber-Gribov  approach \cite{GRIB} where 
\beq \label{E4}
\rho(y_c) \,\,\, \longrightarrow|_{y \gg 1}
\,\,\,a^P_P\,\,A_2 A^{\frac{1}{3}}_1\,\,e^{\Delta\,\, 
Y}  \,\,.  
\eeq

Eq.(~\ref{E4}) predicts a  different behaviour for  the particle density
than does Eq.(~\ref{E3}),  with  a  clear dependence on energy and on $A_1$. 
\eq{E1} also contains  a very simple prediction for the particle density
in the 
fragmentation regions. Assume that $y_c \,\rightarrow 0$
and $y_c \rightarrow\,Y$ correspond to the fragmentation regions for
lightest
and heaviest nucleus, respectively. From \eq{E1} it follows that
\begin{eqnarray}
\rho(y_c)\,\,|_{y_c \rightarrow 0} \,\,\,&=&\,\,A_2\,\,\frac{g_{PN}
\,\gamma}{2 \pi }\,\,; \label{F1}\\
\rho(y_c)\,\,|_{y_c \rightarrow Y}
\,\,&=&\,\,A^{\frac{2}{3}}_2\,\,A^{\frac{1}{3}}_1\,\,\frac{g_{PN}\,\gamma}{2
\pi }\,\,.\label{F2}
\end{eqnarray}
Fig. 2 illustrates  our prediction for particle density.

\begin{figure}[htbp]
\begin{center}
\epsfig{file=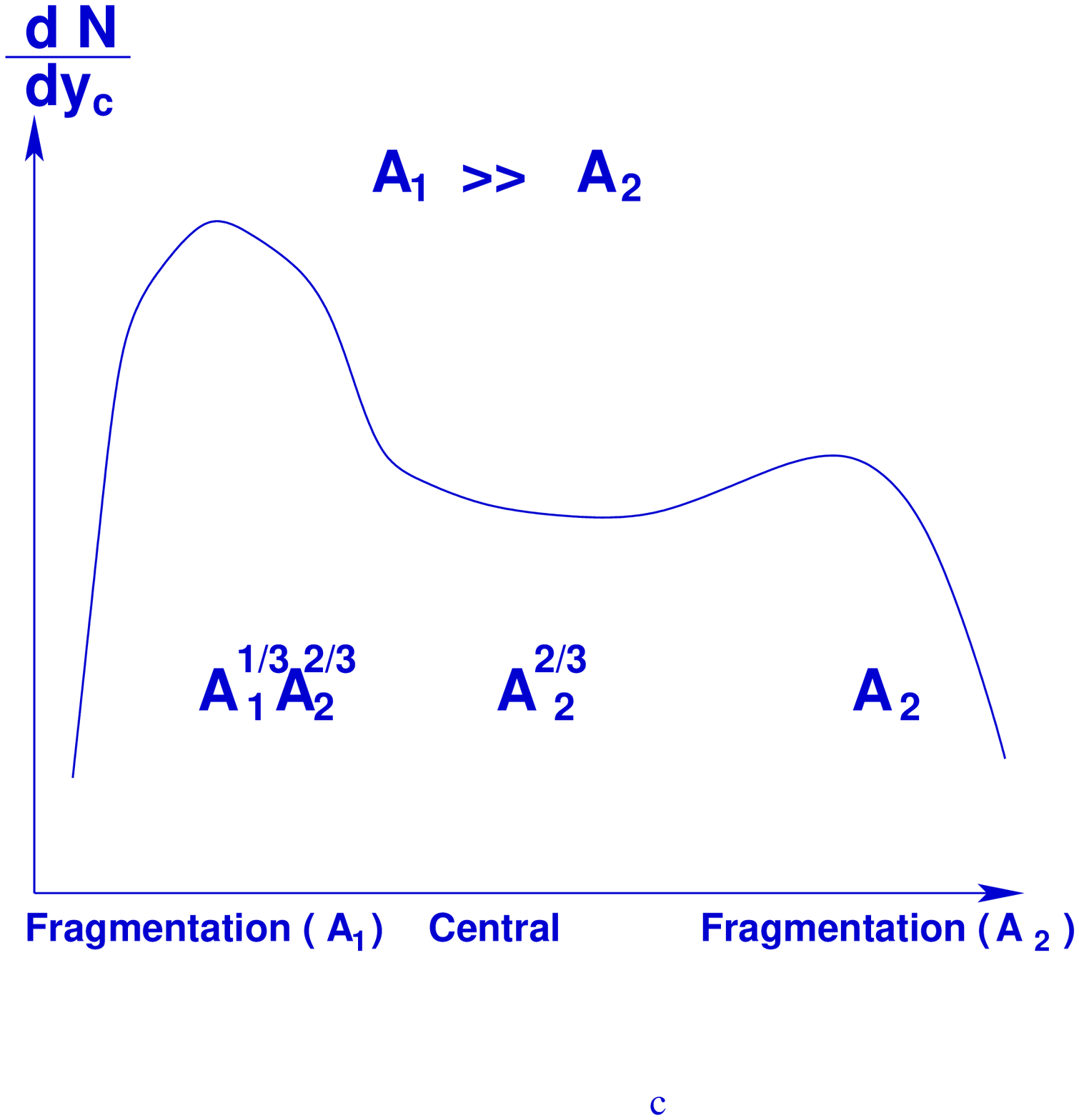,width=10cm}
\end{center}
\caption{\it Our prediction for particle production.}
\label{fig2}
\end{figure}

  The physics  that \eq{E3} is  based on,  is the same saturation of the
parton
density that has been discussed over the past decade in the framework
of high density QCD \cite{GLR,MUQI,MLVE}.  Namely, due to the emission of
partons,  the parton density increases at high energies but,
simultaneously, 
the probability of the interaction of two partons increases as well.
Since,
two interacting partons can annihilate into  one, such interactions 
diminish  the number of partons in the parton cascade. Finally, 
competition between the emission and the annihilation
leads to an equilibrium state with a definite
value for the parton density. We call this phenomenon,   saturation of
the parton density. 
The high density QCD approach leads 
to the  following $A-A$ dependence of
the process \cite{GLR,MUQI,MLVE}:
$$
<N_{A_{1}\,\,A_{2}}> = 
Q_{s}^{2}\Le s, A_1\Ra\cdot
A_{2}^{\frac{2}{3}}\cdot < N_{N\,\,N}>,
$$
here $A_2<A_1$,
$Q_{s}^{2}\Le s, A_1\Ra$ is the saturation scale, which increases with
$A$, and the energy squared $s$,
$<N>$ is the density of the produced hadrons in the 
$A_1-A_2$ or the $N-N$ interactions. $Q_{s}^{2}\Le s, A\Ra$ 
grows with $A$ ( at least $Q_{s}^{2}\propto A^{\frac{1}{3}}$ 
\cite{THEHDQCD}, but it could even be proportional to 
$A^{\frac{2}{3}}$ \cite{KOV} ), and therefore high density
QCD predicts 
\beq \label{HDE}
< N_{A_{1}\,\,A_{2}}>\propto
A_{2}^{\frac{2}{3}}\cdot\Le A_{1}^{\frac{1}{3}}\div 
A_{1}^{\frac{2}{3}}\Ra ,
\eeq 
where $A_{2} < A_{1}$.
\eq{E3} can be rewritten as:
\beq \label{FF1}
<N_{A_{1}\,\,A_{2}}>= 
A_{2}^{\frac{2}{3}}< N_{N\,\,N}>,
\eeq
where $A_{2} < A_{1}$.

 \eq{FF1} leads to  different predictions than those of 
both the Glauber approach (see \eq{E4} ) and the  high density QCD
approach ( see \eq{HDE} ).

 To calculate the particle density at RHIC energies we need to specify how
to calculate the total cross section. It turns out that even in our
approach  for nucleus-nucleus scattering, the problem of determining 
the total cross
section is rather complicated,  but it has been solved in Ref. \cite{AA}.
We use the following expression for $\sigma_{tot}$:
\begin{eqnarray}\label{ST1}
\sigma_{tot}\,\,&=&\,\,2\,\int\,d^2\,b_t\,\,\left(\,1\,\,-\,\,
\exp\Le
-\,\frac{\Omega(s,b_t)}{2}\Ra\,\right)\,\,.\label{O1}\\
\end{eqnarray}
Here $\Omega(s,b_t)$ the  opacity  is given by   
\beq \label{AGA10}
\Omega(s,b_t)\,\,=\,\,\int \,d^2 b'_t \,\,\,F \Le Y, b'_t, | \vec{b}_t -
\vec{b'}_t|\Ra \,\,,
\eeq
and   $F \Le Y, b'_t, | \vec{b}_t -\vec{b'}_t|\Ra$ is the amplitude of
the   nucleus-nucleus interaction \cite{AA}. Since the RHIC energies are
not
very high ( $ Y = \ln( s/1GeV^2) \approx \,10 $ we can use a simple
formula
for $F$ ( see Ref. \cite{AA} for details ):
\begin{eqnarray}
F\Le Y, b_t, b'_t \Ra\,&=&\,\frac{g_{PA_1}(b_t - b'_t)\cdot
g_{PA_2}( b'_t)\cdot e^{\Delta Y}}
{\Le g_{PA_1}(b_t - b'_t) +g_{PA_2}(b'_t) \Ra \gamma\Le e^{\Delta Y}
-1\Ra +1}\,\,.
\label{AGA12}
\end{eqnarray}

 We assume the  Wood-Saxon
parametrization
\cite{WS} for profile function $ S_A(b_t)$.
 
\beq \label{WS}
S_A(b_t)\,\,\,=\,\,\,\frac{\rho}{1\,\,+\,\,e^{\frac{r - R_A}{h}}}\,\,;
\eeq
where $R_A$ is the nucleus radius $ R_A = 1.4 A^{\frac{1}{3}} \,fm$ and 
$h$ the surface thickness $h \approx 1 fm$. For a numerical estimate we
used  the
parameters of \eq{WS} given in  Ref. \cite{WSPAR}.

 It is even easier to calculate the particle density for
central A-A collisions which has  been  measured  at RICH. The 
observable can be written as a ratio of the inclusive cross section
to the cross section at fixed $b_t=0$.

\beq 
\rho_{central}\Le y_c\Ra\,=\,
\frac{\int \,\,\,\,d^2 b'_t
\,\,a^P_P\,\,R_{A_1}\Le Y - y_c, b'_t \Ra \,\cdot\,R_{A_2 }\Le y_c, b'_t
\Ra }{2\,\,\left(\,1\,\,-\,\,
\exp\Le-\,\frac{\Omega(s,b_t=0)}{2}\Ra\,\right)}\,\,\rightarrow
\label{Add2}
\eeq
\beq
\rightarrow\,\,\int \,\,\,\,d^2 b'_t\,\,
\frac{g_{PA_1}(b'_t)\cdot
g_{PA_2}( b'_t)\cdot e^{\Delta Y}}
{\Le g_{PA_1}(b'_t)\gamma\Le e^{\Delta\Le Y-y_c\Ra}
-1\Ra +1\Ra\,\Le g_{PA_2}(b'_t)\gamma\Le e^{\Delta\,y_c}
-1\Ra +1\Ra}
\label{Add1} 
\eeq
since $\Omega(s,b_t=0)\gg\,1$ for A-A scattering.

 We calculate the particle density, defined be \eq{E2},
for Au-Mo, Au-Ne and Ne-Mo scattering at the RHIC energies, 
$W=56\,\,GeV$ ( $Y=\ln\Le s/1\,GeV^2\Ra =8$ ) and
$W=156\,\,GeV$ ( $Y=\ln\Le s/1\,GeV^2\Ra =10$) . 
For proton-proton scattering  we use the following
parameters, which fit the experimental data quite well ( see Ref.
\cite{AA}  for details ):
\begin{center}
\begin{tabular}{l l l l}
$\gamma = 1.19 \,\,GeV^{-1}$ & $ g_{PN} = 8.4 \,GeV^{-1}$ & $ \Delta =
0.07
$ & $a^P_P = 2$.\\
\end{tabular}
\end{center}
Fig. 3 shows that we find a
particle density 
which is $2\div 3$ times smaller than in the  Glauber approach.
The contrast is even more impressive when compared with the prediction 
of the gluon saturation approaches or/and for quark-gluon plasma production
( see Refs.\cite{15} and \cite{16} for example ). The physics is very
simple and is
the same as for  the gluon saturation approach:
the interactions between partons diminish their number, but in contrast
to  gluon saturation the interaction only occurs when  the partons have  
low  transverse 
momenta.

\begin{figure}[htbp]
\begin{tabular}{c c}
{\Large $\mathbf{W\,=\,56\,\, GeV}$} & {\Large $\mathbf{ W\,=\,156\,\,
GeV}$ }\\
 & \\
\epsfig{file=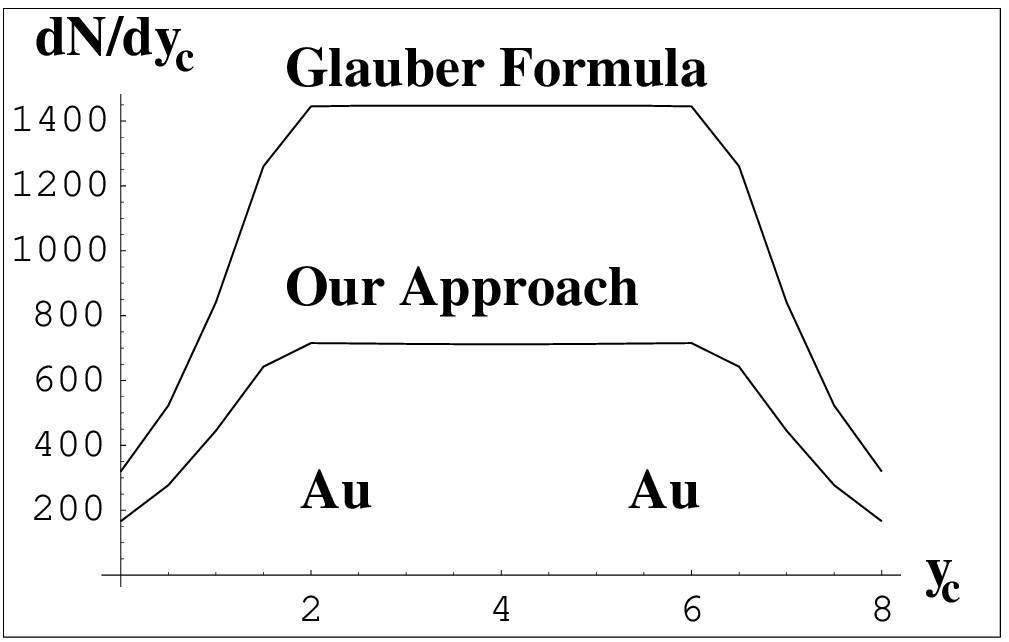,width=8cm,height=4cm} &
\epsfig{file=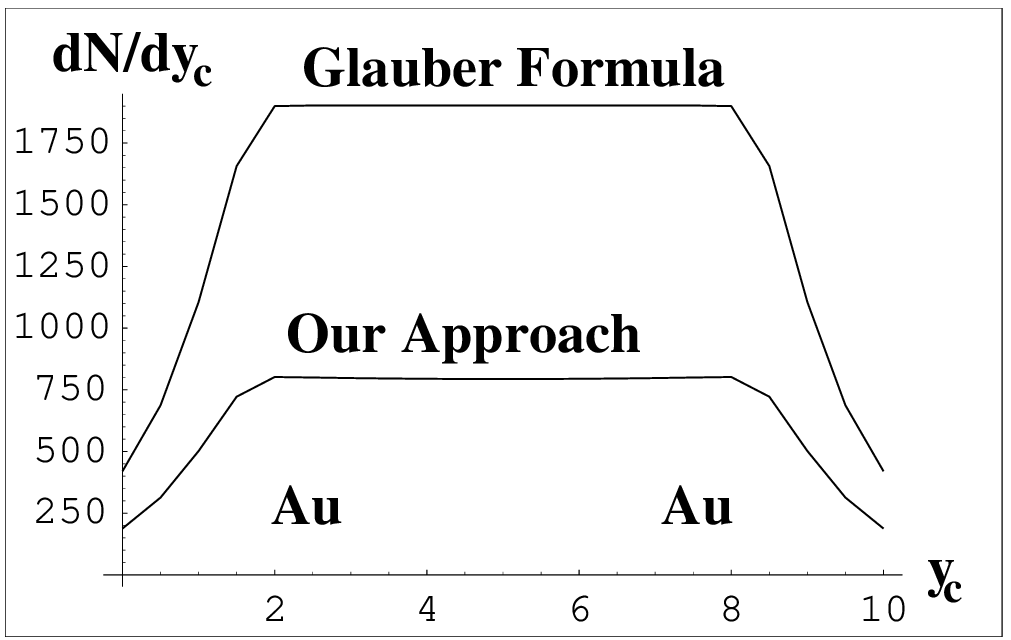,width=8cm,height=4cm} \\
\epsfig{file=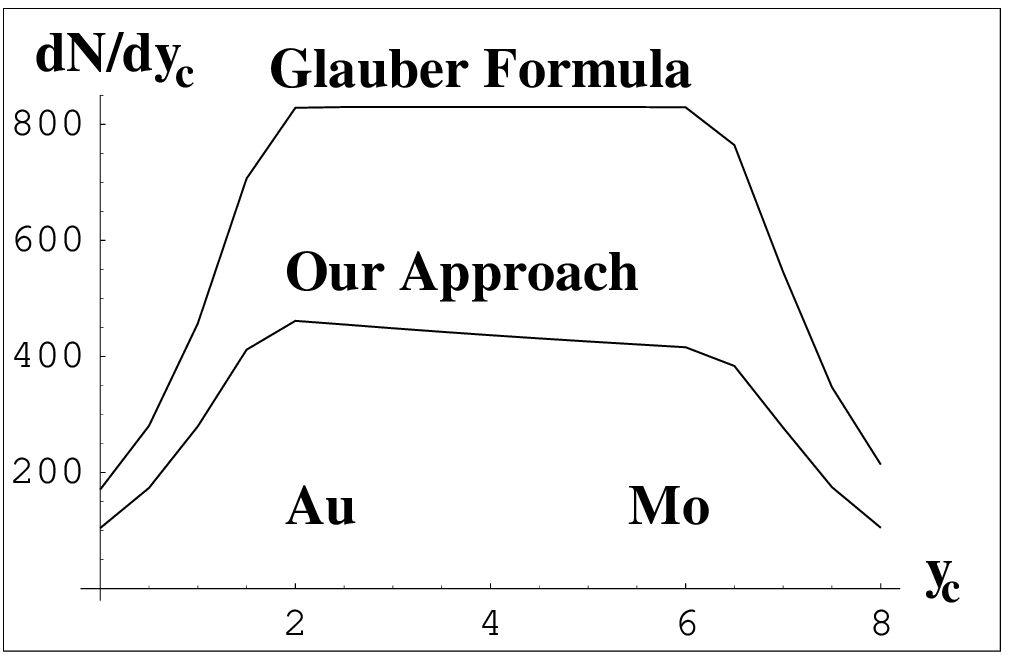,width=8cm,height=4cm} &
\epsfig{file=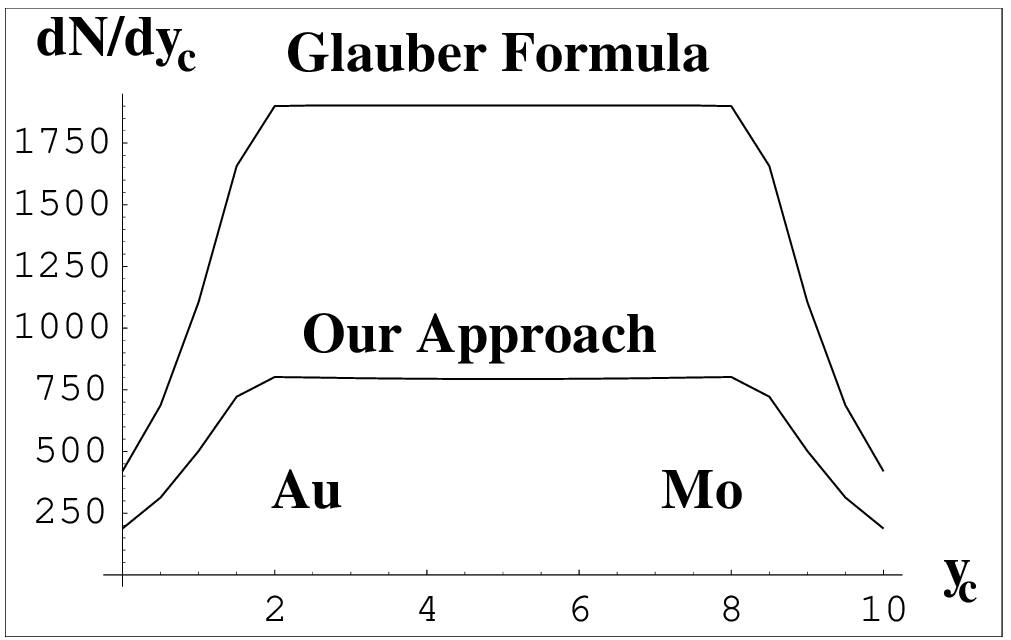,width=8cm,height=4cm} \\
\epsfig{file=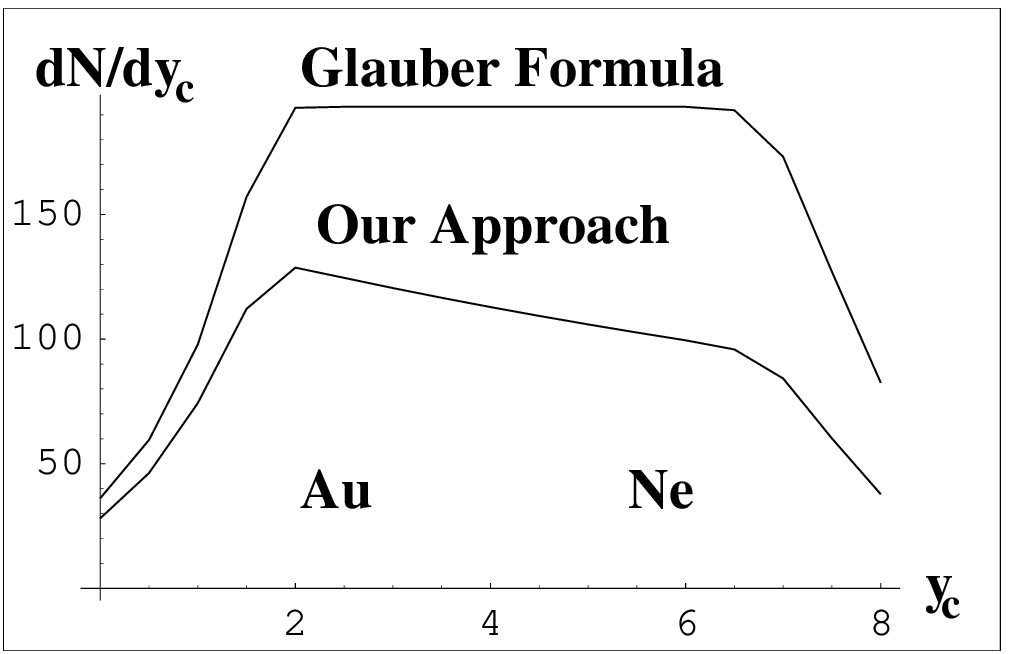,width=8cm,height=4cm} &
\epsfig{file=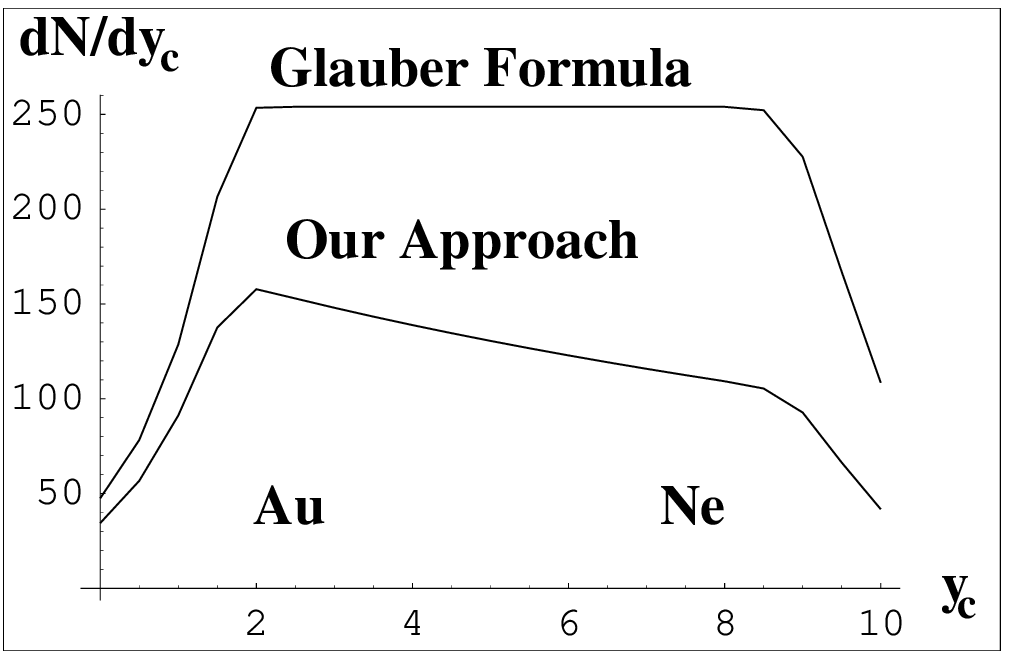,width=8cm,height=4cm} \\
\epsfig{file=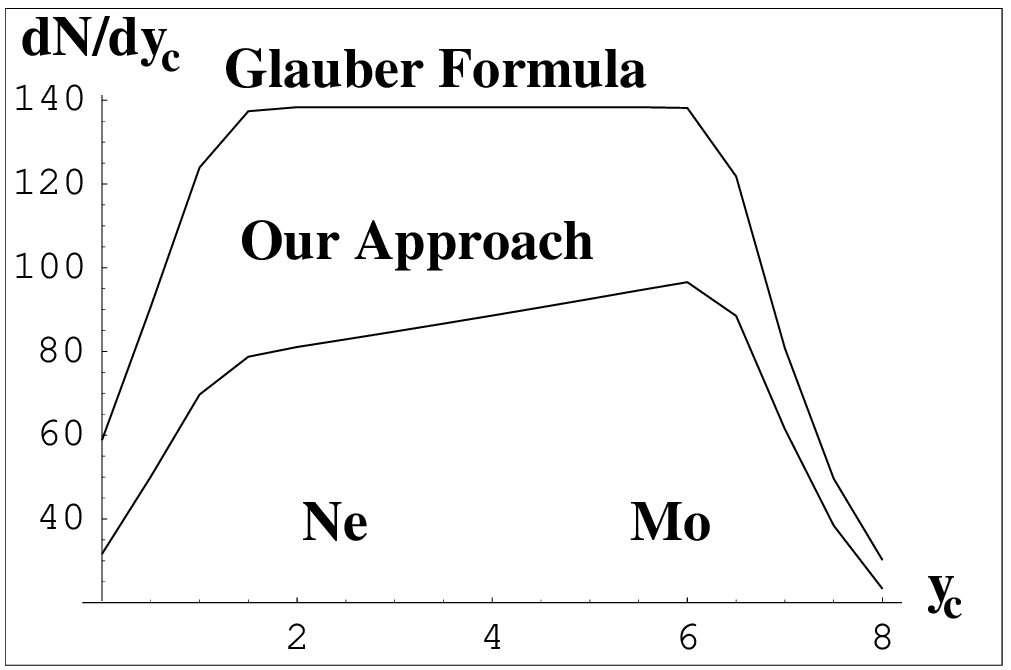,width= 8cm,height=4cm} &
\epsfig{file=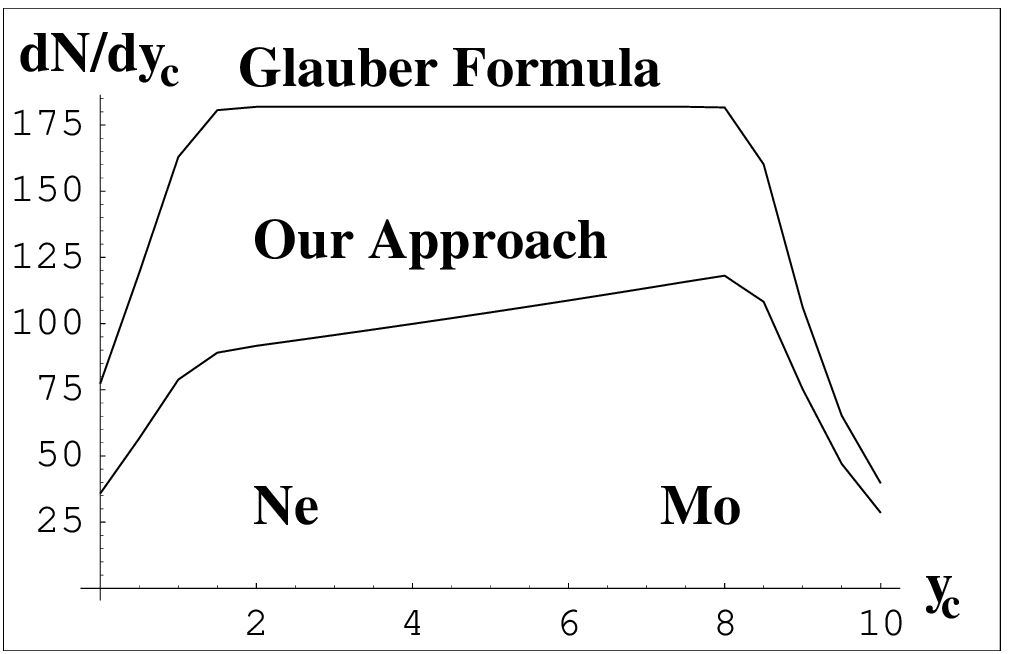,width=8cm,height=4cm} \\
\end{tabular}
\caption{\it The density of produced hadrons for Au-Au, Au-Mo, Au-Ne and 
Mo-Ne in central A-A interactions at RHIC energies.}
\label{fig3}
\end{figure} 
\newpage

Comparison with the experimental data\cite{17}\footnote{ The
experimental data at lowest energy ( $\sqrt{s} = 24\,\,GeV$ ) were
recalculated from lead - lead collision to gold-gold one , using our
formulae.}
  is given in
Fig.4.
We plot three different theoretical curves in Fig.4: the upper one is the
calculation of the hadron density for a  central collision, calculated
using
\eq{Add2} with  $b_t = 0$; the middle curve is calculated ,
assuming that centrality of the nucleus interaction was fixed with an
accuracy $b_t \,\leq \,2\,fm$ \footnote{ We integrated both the numerator
and the denominator in \eq{Add2} over $b_t \leq \,2\,fm$.}; and the lowest 
curve is the inclusive density of \eq{E2}. One can see that the agreement
both in value and in the energy dependence is rather good, but it 
depends strongly on the accuracy with which the centrality of the
experimental
interaction was determined.
However, 
it is very difficult to distinguish our prediction 
  for gold-gold
collisions  from  those based on the saturation approach at high density QCD,
 when the saturation scale is low (approximately $1\div 1.5 \,GeV$ )
  as  was suggested in
Ref. \cite{16}. For 
this case both approaches describe the same physics at RHIC energies and
we need more data on different nuclei to differentiate between   these two
approaches.

 Our conclusions are  simple. We show that the RHIC data
on the rapidity multiplicity in the central A-A collision can be used
to test the different models \cite{GLR,MUQI,MLVE,THEHDQCD,15,16}. We
predict that at RHIC,  densities
will
be rather small ( 400-500 particles  per unit of rapidity for gold-gold
interaction, see Fig.4 ) and these data
 support the saturation hypothesis,  at least for
 sufficiently small transverse momenta for
which our  Pomeron approach is valid. Secondly, our calculations
should be considered as a  background for a signal of new physics
such as saturation of the gluon density in high
energy QCD or/and quark-gluon plasma production. Fig.3 shows that the
measurement of A-dependence will be very decisive in distinguishing
between  the
saturation due to the  Pomeron interaction and the high density QCD
approach.

\begin{figure}[htbp]
\begin{center}  
\epsfig{file=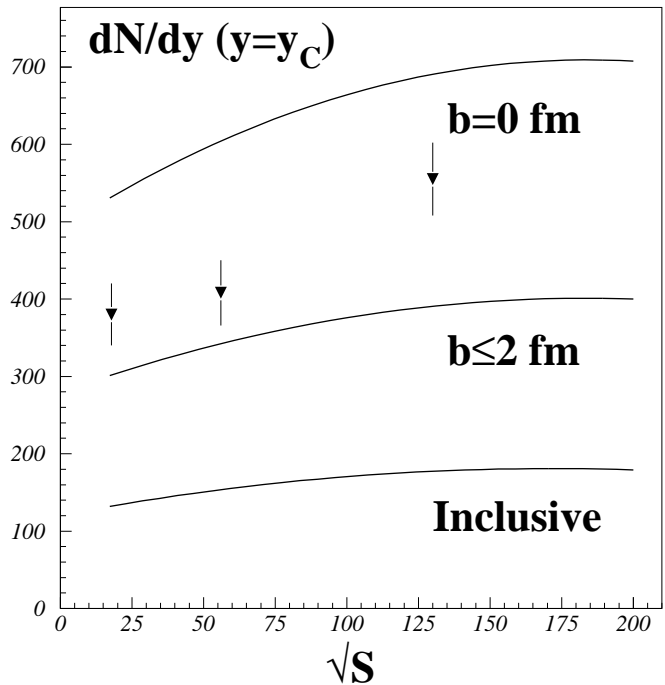,width=10cm}
\end{center}
\caption{\it Energy behaviour of the central hadron density.}
\label{fig4}
\end{figure}

{\bf Acknowledgements:} 
We would like to thank  Larry McLerran, Dima Kharzeev, Yuri Kovchegov and  
Al Mueller, for many
informative and encouraging discussions.

This  research was supported in part by the BSF grant $\#$ 9800276, by the
GIF grant $\#$ I-620-22.14/1999  and by
Israeli Science Foundation, founded by the Israeli Academy of Science
and Humanities.


\begin{thebibliography}{99}
\bibitem{GLR}
L. V. Gribov,
  E. M. Levin and M. G. Ryskin, {\it Phys.Rep.} {\bf 100}, 1 (1983).
\bibitem{MUQI}
A.H. Mueller and J. Qiu, {\it Nucl. Phys.}  {\bf B268}, 427 (1986).
\bibitem{MLVE}
L. McLerran and R. Venugopalan, {\it Phys. Rev.} {\bf D49} (1994)
2233,3352, {\bf D50} (1994) 2225, {\bf D53} (1996) 458, {\bf D59}
(1999) 094002.
\bibitem{Bass}
S.A.Bass et. al.. {\it "Last call for RHIC predictions"},
{\it Nucl.Phys.} {\bf A661}, (1999) 205 and reference therein.
\bibitem{AA}
S.~Bondarenko, E. ~Gotsman, E. ~Levin and U. ~Maor, 
TAUP 2616-99,BNL-NT-99/11,{\tt hep-ph/0001260 } {\it Nucl. Phys. {\bf A}
(in press)}.
\bibitem{PARTONS}
R.P. ~Feynman, {\it Phys. Rev. Lett.} {\bf 23} (1969) 1415,
{\it " Photon-Hadron Interactions"},N.Y. Benjamin, 1972;
J.D. ~Bjorken, {\it Phys. Rev.} {\bf 179} (1969) 1547;
V.N. ~Gribov, { \it Sov.J.Nucl.Phys.} {\bf 9} (1969) 369,
 {\it ``Space-time description of the hadron interaction at high
energies",} Moscow 1 ITEP school, v.1, p.65,\, {\tt 
hep-ph/0006158 }.
\bibitem{SCHWIM}
O. Kancheli and S. Matinian, { \it Sov.J. Nucl.Phys.} {\bf 11}
(1970) 726;
 A. Schwimmer, {\it Nucl. Phys.} {\bf B94} (1975) 445.
\bibitem{AGK}
 V.A. Abramovsky, V.N. Gribov and O.V. Kancheli, {\it Sov. J.
Nucl. Phys.}  {\bf 18} (1974)  308.
\bibitem{JEN}
L. Caneschi, A. Schwimmer and R. Jengo, {\it Nucl. Phys.} {\bf B108}  
(1976) 82.
\bibitem{GRIB}
R.J. Glauber,\, {\it  Lectures in Theoretical Physics.} Ed.\, 
W.E.  Britten
N.Y., Int.Publ.\, 1959,\, v.1,\, \,p.315 ;\,\,\,\,
 V.N. Gribov,{\it JETP}  {\bf 56}(1969)  892, {\bf 57} (1969)  1306.

\bibitem{THEHDQCD}
E.M.   Levin and M.G. Ryskin, {\it Phys. Rept.} {\bf 189} (1990) 267;\\
E. Laenen and E. Levin, {\it Ann. Rev. Nucl. Part.} {\bf 44} (1994) 199
and references therein;\\
A.L. Ayala, M.B. Gay Ducati and E.M. Levin,  {\it Nucl. Phys.}
{\bf B493}, 305 (1997),
{\bf B510}, 355 (1998);\\
Yu. Kovchegov, {\it Phys.Rev.}{\bf D54} (1996) 5463; {\bf D55} (1997)
5445;  {\bf D60} (1999) 034008;\\
A.H. Mueller, {\it  Nucl.Phys.} {\bf B572} (2000)227,
{\it Nucl.Phys.} {\bf B558} (19999) 285;\\
 Yu. V. Kovchegov, A.H. Mueller,   {\it 
 Nucl.Phys.}{\bf B529} (1998)451.
\bibitem{KOV}
Yu. Kovchegov,{\it Phys.Rev.} {\bf D61} (2000) 074018;
E. Levin and K. Tuchin, { \it Nucl.Phys.} {\bf B573} (2000) 833. 
\bibitem{WS}
Harald A. Enge, {\it "Introduction to Nuclear Physics"}, Addison-Wesley
P.C.Inc., 1971.
\bibitem{WSPAR}
De Jager, de Vries and de Vries, {\it "Atomic Data and Nuclear Data
Tables"},
{\bf Vol. 14}, No. 5,6, (Nov/Dec 1974)
\bibitem{15}
N.Armesto and C.Pajares, {\it Int. J. Mod. Phys.} {\bf A15} (2000) 2019, 
{\tt hep-ph/0002163};
A.Krasnitz and R.Venugopalan, {\it "The initial gluon multiplicity in
heavy
ion collision"},{\tt hep-ph/0007108}; {\it "Nonpertubative gluodinamics of
high energy heavy ion collision"}, {\tt hep-ph/0004116};
    {\it Phys.Rev.Lett} {\bf 84} (2000) 4309.
\bibitem{16}
K.J.Eskola,    K. Kajantie, P.V.  Ruuskanen and K. Tuominen,
{\it Nucl.Phys.} {\bf 570} (2000) 379 and 
references therein;\,\, K.J.Eskola,    K. Kajantie and
K. Tuominen,  {\it ``Centrality dependence of multiplicities in
ultrarelativistic nuclear collisions"}, HIP-2000-45/TH, {\tt
hep-ph/0009246}.

 \bibitem{17}
PHOBOS collaboration: B.B. Back et al., {\it Phys. Rev. Lett.} {\bf 85}
(2000) 3100, {\tt hep-ex/0007036}.

\end{thebibliography}
\end{document}